\documentclass[prx,aps,twocolumn,superscriptaddress,bibnotes]{revtex4-1}

\usepackage{dcolumn}
\usepackage{bm}
\usepackage{amssymb}
\usepackage{color}
\usepackage[pdftex]{graphicx}

\usepackage[mathscr]{euscript}

\usepackage{amsmath}
\usepackage[linesnumbered,ruled]{algorithm2e}

\usepackage{url}
\usepackage[section]{placeins}
\usepackage[colorlinks=true,linkcolor=blue,citecolor=blue]{hyperref}
\usepackage{comment}

\begin{document} 
\let\vec\mathbf

\title{Machine learning for phase ordering dynamics of charge density waves}

\author{Chen Cheng}
\thanks{ORCID:https://orcid.org/0000-0003-3224-1231}
\affiliation{Department of Physics, University of Virginia, Charlottesville, VA 22904, USA}

\author{Sheng Zhang}
\affiliation{Department of Physics, University of Virginia, Charlottesville, VA 22904, USA}

\author{Gia-Wei Chern}
\thanks{gchern@virginia.edu }
\affiliation{Department of Physics, University of Virginia, Charlottesville, VA 22904, USA}

\date{\today}

\begin{abstract}
We present a machine learning (ML) framework for large-scale dynamical simulations of charge density wave (CDW) states. The charge modulation in a CDW state is often accompanied by a concomitant structural distortion, and the  adiabatic evolution of a CDW order is governed by the dynamics of the lattice distortion. Calculation of the electronic contribution to the driving forces, however, is computationally very expensive for large systems. Assuming the principle of locality for electron systems, a neural-network model is developed to accurately and efficiently predict local electronic forces with input from neighborhood configurations. Importantly, the ML model  makes possible a linear complexity algorithm for dynamical simulations of CDWs. As a demonstration, we apply our approach to investigate the phase ordering dynamics of the Holstein model, a canonical system of CDW order.  Our large-scale simulations uncover an intriguing growth of the CDW domains that deviates significantly from the expected Allen-Cahn law for phase ordering of Ising-type order parameter field. This anomalous domain-growth could be attributed to the complex structure of domain-walls in this system. Our work highlights the promising potential of ML-based force-field models for dynamical simulations of functional electronic materials.
\end{abstract}

\maketitle

\section{introductin}

\label{sec:intro}

A charge density wave (CDW) is a periodic modulation of electron charge density which breaks translational symmetries of the system~\cite{Gruner1988,Gruner1994, Thorne1996, Monceau2012, Pokrovskii2013}. In many materials, the charge density modulation is accompanied by a structural distortion due to electron-lattice couplings~\cite{Gruner1988,Gruner1994}. Indeed, while CDW order could originate from pure electronic mechanisms~\cite{Tranquada1995,Kivelson2003}, electron-lattice couplings play a crucial role in the emergence of the charge modulation in most CDW materials. For example, the well-known Peierls transition, which describes the instability of a one-dimensional (1D) metal towards the formation of a gapped CDW state, is caused by a periodic lattice distortion that opens a gap at the Fermi points of a partially filled band~\cite{Peierls1955}. The collective nature of charge and lattice dynamics in CDW states leads to several interesting phenomena such as nonlinear electrical conduction, giant dielectric response, and multi-stable conducting states, to name a few~\cite{Monceau2012, Anderson1973,Balandin2021}. A reinvigorated interest in the CDW physics was recently sparked by the synthesis of quasi-two-dimensional (2D) materials such as transition metal dichalcogenides~\cite{Porer2014, Stojchevska2014, Samnakay2015, Cho2016, Vogelgesang2018, Hellmann2012}.

The research on CDW phases has largely focused on their thermodynamic, electronic, and transport properties, as well as their interplay with other symmetry-breaking phases, such as superconductivity.  A new frontier of research, enabled by the advent of ultrafast technology, is the formation  and coherent dynamics of CDW orders. Such experiments aim to study the electronic and structural dynamics of CDW materials under short-pulse excitations with femtoseconds or picoseconds resolution~\cite{Maklar2021, Yusupov2010, Schaefer2014, Trigo2019, Dolgirev2020}. Despite enormous interest in the ultrafast dynamics of CDW systems, their dynamical behaviors in the adiabatic limit, such as the phase-ordering kinetics, have yet to be systematically investigated, especially on the theoretical side.  Important issues such as whether the coarsening of CDW domains follows well-established power laws or exhibits dynamical scaling remain open. Understanding the adiabatic dynamics will not only serve as important references for the ultrafast dynamics, but also shed light on the morphology of CDW states, especially in the presence of impurities and strains~\cite{tsen15,gao18,sharma20}.

A full quantum dynamical simulation of the coarsening dynamics of CDW is, however, a difficult multi-scale task. On the one hand, large-scale systems are required in order to capture details of pattern formation processes at the mesoscale. On the other hand, in order to accurately model the electronic driving forces, one needs to carry out time-consuming quantum computations, ranging from exact diagonalization to more sophisticated many-body techniques, at every time step of the dynamical simulation. Indeed, in most large-scale simulations of CDW phenomena to date, the electronic degrees of freedom are modeled by an order-parameter field governed by the phenomenological time-dependent Ginzburg-Landau (TDGL) equation. More accurate quantum approaches to the electronic structure calculation of CDW dynamics, however, are often restricted to small systems or simply neglect spatial fluctuations or inhomogeneity.

In this paper, we present a solution to the multi-scale modeling of adiabatic CDW dynamics based on a scalable machine learning (ML) framework. Our approach utilizes the so-called nearsightedness of electronic matter~\cite{Kohn1996, Prodan2005}, which assumes fast decays of electron correlation functions. This in turn implies that local electronic properties depend predominantly on the immediate environment.  In the case of adiabatic CDW dynamics, the locality principle implies that the electronic driving forces acting on  local lattice degrees of freedom of a CDW order can be accurately determined from structural information in the neighborhood. A deep-learning neural network model is then developed to encode this complicated dependence on the neighborhood configurations. Importantly, our proposed ML force-field model is both transferrable and scalable, which means that models trained from dataset of small-scale exact solutions can be directly applied to much larger systems. 

It is worth noting that our ML approach to the adiabatic CDW dynamics is similar in spirit to ML-based interatomic potential models that are used to enable large-scale {\em ab initio} molecular dynamics (MD) simulations. In such first-principles MD methods, the atomic forces are obtained by solving e.g. the Khon-Sham equation at every time-step~\cite{Marx}.  ML models are developed to accurately emulate the time-consuming electronic structural calculations. Again, the locality principle is implicitly assumed in such linear-scaling ML-based quantum MD methods. A similar ML framework has recently been proposed for multi-scale dynamical modeling of general condensed matter systems~\cite{zhang22a}. Large-scale simulations based on such scalable ML models have also been demonstrated in itinerant magnets and strongly correlated electron models~\cite{ma19,liu22,zhang22b,zhang20,zhang21a,zhang21b}.

In this work we apply our ML model to investigate the coarsening of CDW domains in the Holstein model~\cite{Holstein1959}, a canonical system for CDW physics~\cite{noack91,zhang19,chen19,hohenadler19}. The Holstein model is a lattice model of itinerant electrons interacting with scalar dynamical variables, which represent local $A_1$-type structural distortions associated with each lattice sites. For bipartite lattices in both 2D and 3D, the half-filled electron band is unstable towards the formation of a checkerboard charge-density modulation, which is accompanied by a staggered arrangement of local lattice distortions. The CDW phase transitions and other phenomena related to electron-phonon coupling, such as polaron dynamics and superconductivity, have been extensively studied in the Holstein model~\cite{bonca99,golez12,mishchenko14,scalettar89,costa18,bradley21}. Yet, to the best of our knowledge, the fundamental phase-ordering dynamics of the CDW order, even in the semiclassical approximation, has never been investigated. 

The study of phase-ordering dynamics concerns the growth and coarsening of ordered domains when a system is quenched into a symmetry-breaking phase~\cite{Bray1994, Puri2009, Onuki2002, Cross2012}. The evolution of the order-parameter fields is often highly nonlinear and is characterized by the emergence of complex spatial patterns. The difficulty of large-scale coarsening simulations of the Holstein model, as alluded to previously, is due to the multi-scale nature of the problem. With the aid of ML force-field model trained from small-scale exact solutions, we performed phase-ordering simulations of the Holstein model subject to a thermal quench on large systems of $\sim 10^5$ sites. Our results show that the phase-ordering of CDW exhibits dynamical scaling and power-law domain growth. Yet, the growth exponents are different from values that are expected from symmetry consideration and conservation laws. Moreover, the exponents exhibit intriguing dependences on temperature and electron filling-fraction, highlighting the nontrivial interplay between the electron and the lattice degrees of freedom.

The rest of the article is organized as follows. The Holstein model and its adiabatic dynamics are discussed in Sec.~\ref{sec:Holstein}. We outline the general ML framework for adiabatic CDW dynamics, and present a specific scalable neural-network model for the dynamical simulation of the Holstein model  in Sec.~\ref{sec:ML}. Benchmarks of force prediction and comparison of dynamical simulations on small-scale systems are also discussed. Results of large-scale ML-enabled phase-ordering simulations and analysis of the growth dynamics of the CDW order are discussed in Sec.~\ref{sec:result}. Finally, Sec.~\ref{sec:summation} concludes the article with a summary and outlook.

\section{Adiabatic dynamics of the Holstein model} 
\label{sec:Holstein}

While our ML approach can in principle be applied to adiabatic dynamics of general CDW orders, as a proof-of-principle, and for concreteness, here we demonstrate the method using the Holstein model~\cite{Holstein1959}. We consider a modified Holstein model with spinless fermions on a square lattice:
\begin{align}
	& &  {\mathscr{H}} = -t_{\rm nn} \sum_{\langle ij \rangle}  {c}_i^\dagger  {c}_j - g \sum_i \left( c^\dagger_i c^{\,}_i - \frac{1}{2} \right) Q_i \notag\\
	& & \qquad  +\sum_i \left( \frac{P_i^2}{2 m} + \frac{k Q_i^2}{2} \right)+ \kappa \sum_{\langle ij \rangle} Q_i Q_j.
	\label{eq:H_holstein}
\end{align}
Here $ {c}^\dagger_i$ ($ {c}_i$) is the electron creation (annihilation) operator at site-$i$, $Q_i$ denotes a scalar dynamical lattice degree of freedom associated at the $i$-th lattice site, and $P_i$ is the corresponding conjugate momentum. The first term above describes the electron hopping between a pair of nearest-neighbor sites $\langle ij \rangle$, with $t_{\rm nn}$ being the nearest-neighbor hopping coefficient. The second term represents a deformation-type electron-lattice coupling, where $g$ is the coupling constant and ${c}^\dagger_i  {c}^{\,}_i$ is the electron number operator. The lattice degrees of freedom here are modeled by a set of simple harmonic oscillators, with effective mass $m$ and elastic or spring constant $k$. Finally, the last term introduces a quadratic interaction $\kappa$ between nearest-neighbor harmonic oscillators. 

The lattice degrees of freedom $\{Q_i\}$ in the Holstein model are similar to the Einstein model of dispersionless phonons. The model could also be used to describe real compounds, where $Q_i$ represent amplitudes of local collective modes of atomic clusters such as the breathing mode of an octahedron centered at site-$i$. Partly due to its relative simplicity, the Holstein model and its variants are amenable to quantum Monte Carlo (QMC) methods, and are widely used as a minimum model to study the physics of electron-phonon coupling, such as polaron dynamics and phonon-mediated superconductivity (SC). At half-filling on a bipartite lattice, including both square and honeycomb, the model exhibits a transition to the CDW order at a finite temperature~\cite{noack91,zhang19,chen19,hohenadler19}. As the system is doped away from half-filling, SC correlation is enhanced and quasi-long-range SC order eventually sets in at very low temperatures~\cite{bonca99,golez12,mishchenko14,scalettar89,costa18,bradley21}.

The CDW order of half-filled Holstein model on bipartite lattices is characterized by a checkerboard electron density modulation: $n_{A/B} = (1 \pm \delta)/2$, where the subscript $A$ and $B$ refers to the two sublattices of the bipartite lattice, and $\delta$ quantifies the charge modulation. Due to the electron-lattice coupling, the checkerboard charge modulation is accompanied by a staggered lattice distortion $Q_{A/B} = \pm \mathcal{Q}$. The CDW order of the Holstein model thus breaks the $Z_2$ sublattice symmetry, which is a special commensurate translational symmetry breaking. On symmetry ground, the CDW transition is expected to belong to the Ising universality class, which is indeed confirmed by QMC simulations~\cite{noack91,zhang19}. The CDW phase of a half-filled Holstein model is thus described by an Ising-type order parameter field~\cite{mcmillan75}.

As numerous other physical systems also exhibit an Ising-type phase transition, the ordering dynamics of Ising phases is one of the most studied subjects and has been successfully used to describe coarsening phenomena in many materials ranging from magnets with an easy-axis anisotropy to binary alloys. The phase ordering behaviors of Ising systems have been thoroughly characterized and classified into several super-universal classes which depend on whether the Ising order is conserved and the presence of quenched disorder. Despite extensive studies on the equilibrium properties of Holstein models, it remains unclear whether the transition dynamics of the CDW order is consistent with these universal behaviors.

A full quantum treatment of phase transition dynamics is extremely difficult, even for Holstein models. Since the system remains out of equilibrium during phase ordering, the powerful QMC methods cannot be applied in such dynamical studies. To make numerical simulations tractable, here we introduce the semiclassical approximation and treat the lattice degrees of freedom as classical dynamical variables. Unlike the SC phase which requires full quantum treatment, the CDW order remains robust even in the semiclassical approximation. Indeed, the semiclassical phase diagram of the CDW order obtained by a hybrid Monte Carlo method agrees very well with that obtained from DQMC simulations~\cite{esterlis19}.

Within the semiclassical approximation, the dynamics of the local lattice modes is described by the effective Newton equation of motion
\begin{eqnarray}
	\label{eq:langevin}
	m\frac{d^2Q_i}{dt^2} = - \frac{\partial \langle  \mathscr{H} \rangle}{\partial Q_i}  - \gamma \frac{dQ_i}{dt} + \eta_i(t)
\end{eqnarray}
Here the Langevin thermostat is used to account for the effects of a thermal reservoir during the phase ordering; $\gamma$ is a damping constant and $\eta_i(t)$ is a thermal noise of zero mean described by correlation functions 
\begin{eqnarray}
	 \langle \eta_i(t) \rangle &=& 0, \\
	 \langle \eta_i(t) \eta_j(t') \rangle &=& 2 \gamma k_B T \delta_{ij} \delta(t - t'). \nonumber
\end{eqnarray}
Central to the integration of the above Langevin equation is the force calculation which requires the computation of the expectation value of the Hamiltonian $\langle \mathcal{H} \rangle$. With the semiclassical approximation, the force term can be separated into the contribution from electrons and a classical elastic force
\begin{eqnarray}
	\label{eq:force-total}
	& & F_i \equiv - \frac{\partial \langle \mathscr{H} \rangle}{\partial Q_i} = F^{\rm elastic}_i + F^{\rm elec}_i \nonumber \\
	& & \qquad \qquad = -\frac{\partial \mathcal{V}}{\partial Q_i} - \frac{\partial \langle \mathcal{H}_e \rangle}{\partial Q_i}. 
\end{eqnarray}
Here the classical potential is $\mathcal{V}(\{Q_i\}) = \frac{k}{2} \sum_i Q_i^2 + \kappa \sum_{\langle ij \rangle} Q_i Q_j$, and the resultant force is simply the restoring force of the simple harmonic oscillator with an additional coupling term
\begin{eqnarray}
	\label{eq:F_elastic}
	F^{\rm elastic}_i = - k Q_i - \kappa \sum_{j}\phantom{}^{'}  Q_j,
\end{eqnarray}
where the prime in the second term indicates the summation is restricted to nearest neighbors of site-$i$. The electron Hamiltonian in Eq.~(\ref{eq:force-total}) corresponds to a tight-binding model with a random on-site chemical potential due to coupling to lattice distortions
\begin{eqnarray}
	\label{eq:H_elec}
	\mathcal{H}_e(\{Q_i\}) = -t_{\rm nn} \sum_{\langle ij \rangle}  {c}_i^\dagger  {c}_j - g \sum_i \left( c^\dagger_i c^{\,}_i - \frac{1}{2} \right) Q_i.
\end{eqnarray}
The calculation of the electronic force can be simplified using the Hellmann-Feynman theorem~\cite{Ventra2000, Todorov2001, Lu2012, Todorov2010, Dundas2009} $\partial \langle \mathcal{H}_e \rangle / \partial Q_i = \langle \partial \mathcal{H}_e / \partial Q_i \rangle$, which gives 
\begin{eqnarray}
	\label{eq:F_elec}
	F^{\rm elec}_i = g \left(  n_i  - \frac{1}{2} \right),
\end{eqnarray}
where $n_i \equiv \langle c^\dagger_i c^{\,}_i \rangle$ is the expectation value of the electron number. 
The electron force is proportional to the deviation of the average local electron density from half-filling.

The domain growth during phase ordering is in general a slow process compared with the relaxation of electrons. The evolution of the CDW state can thus be well described by the adiabatic approximation. Specifically, we assume the time-scales of the slower lattice dynamics is well separated from the fast electron relaxation. It is worth noting that this adiabatic approximation is exactly the same as the Born-Oppenheimer approximation widely used in quantum MD simulations~\cite{Marx}. The fast electron relaxation indicates that the electronic subsystem reaches quasi-equilibrium with respect to the instantaneous lattice configuration. The expectation value of, e.g. the electron density, is thus computed from a Boltzmann distribution
\begin{eqnarray}
	 n_i  = \frac{1}{Z_e} {\rm Tr}\left( c^\dagger_i c^{\,}_i  \, e^{-\beta \mathcal{H}_e(\{Q_i\})} \right),
\end{eqnarray}
where $Z_e = {\rm Tr}e^{-\beta \mathcal{H}_e(\{Q_i\})}$ is the partition function of quasi-equilibrium electrons. However, even with the adiabatic approximation, the dynamical evolution of the Holstein model is still a computationally demanding task. As the electron Hamiltonian $\mathcal{H}_e$ is quadratic in the fermion creation/annihilation operators, it can be solved by the exact diagonalization (ED) in real-space. Yet, since the electronic forces have to be computed at every time-step of the Langevin dynamics simulation, the $\mathcal{O}(N^3)$ time complexity of  ED can be overwhelmingly time-consuming for large systems.

The ML framework to be discussed in the next section, on the other hand, provides a linear complexity algorithm for the force calculation, which is key to large-scale phase-ordering simulations of the CDW order.  We also note that other linear-scaling numerical techniques, notably the kernel polynomial methods~\cite{weisse06,wang18}, have been developed to solve quadratic fermion Hamiltonians. However, these other $\mathcal{O}(N)$ methods cannot be directly generalized to fermionic models with electron-electron interactions such as the on-site Hubbard repulsion. In this regard, the ML methods offer a general approach to achieve linear scalability even for strongly correlated systems such as the Holstein-Hubbard model~\cite{berger95,johnston13,costa20}.

\section{Machine learning force-field model and benchmarks} 

\label{sec:ML}

As discussed in Sec.~\ref{sec:intro}, the feasibility of linear-scaling electronic structure methods is fundamentally due to the assumption of locality, or nearsightedness, of many-electron systems~\cite{Kohn1996,Prodan2005}. Modern ML techniques provide an explicit and efficient approach to incorporate the locality principle into the implementation of $\mathcal{O}(N)$ methods. 
Perhaps one of the most prominent applications in this regard is the ML-based interatomic potential or force-field models for {\em ab initio} MD simulations~\cite{Behler2007,Bartok2010, Li2015, Smith2017, Zhang2018, Behler2016, Deringer2019, Mueller2020, McGibbon2017, Chmiela2017, Suwa2019,chmiela18,sauceda20}. Unlike classical MD methods where empirical formulas are used for the force calculation, a many-electron Schr\"odinger equation is numerically solved, with varying approximations, at every time-step in order to compute the atomic forces in quantum MD approaches~\cite{Marx}. Various ML models have been proposed to emulate the time-consuming electronic structure calculations. Again, locality is tacitly assumed in such ML-MD methods, which means that the atomic force only depends on the immediate surrounding of the atom under consideration. 

Generally speaking, there are two different approaches to the force calculation in the ML-based quantum MD methods. In the energy-based models, first proposed by Behler and Barrinello (BP)~\cite{Behler2007} and by Bart\'ok {\em et al.}~\cite{Bartok2010}, the total energy of the system is partitioned into local atomic contributions: $E = \sum_i \epsilon_i$. Assuming locality of these atomic energies, ML models are developed to accurately approximate their dependence on the neighborhood atomic configurations. The atomic forces are then computed from the derivatives of the energy $\mathbf F_i = -\partial E / \partial \mathbf R_i$. One immediate advantage of this approach is that the resultant force is conservative, which is important for quantum MD within the Born-Oppenheimer approximation. Moreover, by focusing on the energy which is a scalar invariant under the rotation group, symmetry constraints can be more easily incorporated into such ML models. 

In the second approach, ML models are used to directly approximate the global force fields. The most representative examples of this approach is the gradient-domain machine learning (GDML) models~\cite{Chmiela2017,chmiela18,sauceda20}.  To ensure energy conservation, a selected set of vector-valued kernel functions are used to approximate the force fields. As the vector-output is not invariant under symmetry transformations, special care has to be taken to preserve symmetries of the molecular systems.  The GDML and its variants have been shown to achieve high accuracy force prediction with relatively small-size of training dataset.

\begin{figure*}	
\centering
\includegraphics[width=1.99\columnwidth]{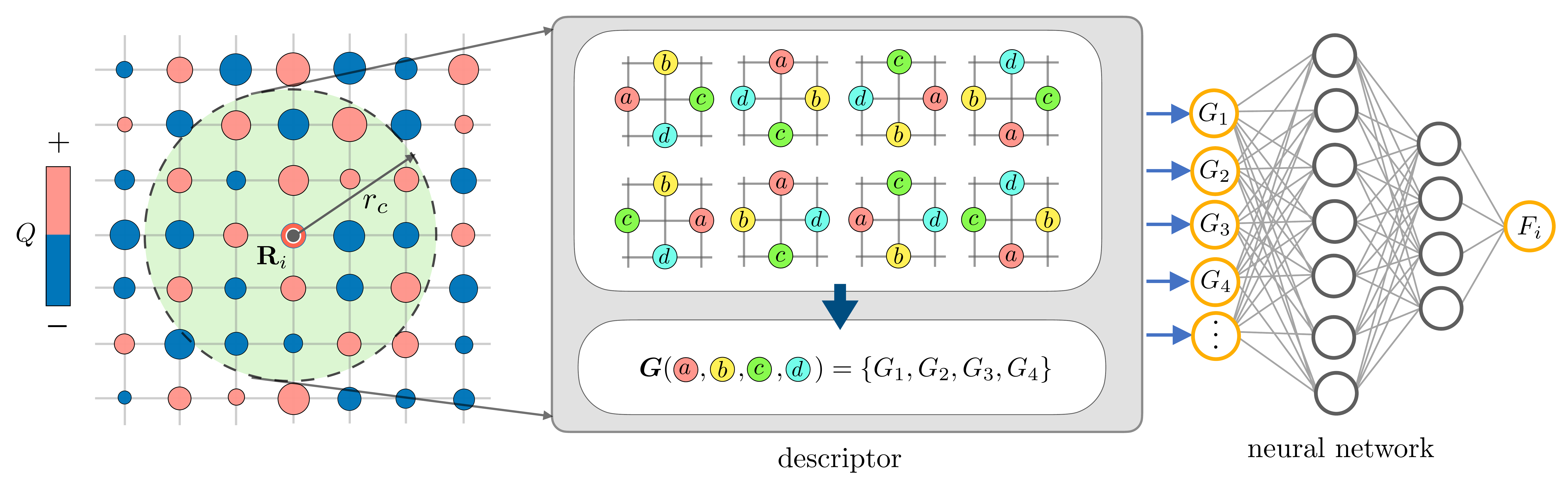}
\caption{Schematic diagram of the ML force-field model for the Holstein model. The input of the ML model is the lattice configuration $\mathcal{C}_i$ centered at site-$i$,  while the output is the force $F_i$ acting on the associated lattice distortion. There are two central components of the ML model: (i) the descriptor and (ii) a multi-layer neural network. The lattice descriptor is introduced to preserve the $D_4$ point-group symmetry of the model. Essentially 8 symmetry-related configurations are mapped to the same feature variables $\{ G_\ell \}$ which are then fed into the neural network.}
\label{fig:NN}
\end{figure*}

The relative simplicity of the lattice degrees of freedom in the Holstein model suggests a direct force-field ML model. Indeed, since the effective force acting on local lattice mode $Q_i$ is a scalar, it is already invariant under symmetry transformations of the system, in contrast to vector forces in MD simulations. By constructing neighborhood feature variables that are also invariants of the symmetry group, the symmetry of the Holstein model can be readily incorporated into the ML model. Finally, since forces in energy-based ML models have to be obtained through derivatives of the total energy, additional overhead is required for this calculation. Consequently, both the training and the performance of the direct-force ML model are computationally more efficient. 

As the elastic force Eq.~(\ref{eq:F_elastic}) can be trivially obtained analytically, one is tempted to focus on the modeling of the electronic forces in Eq.~(\ref{eq:F_elec}), which is proportional to the local electron number $ n_i $. However, when the system approaches an insulating CDW state, the on-site electron density tends to $n \sim 1$ or $0$. The electronic force thus exhibits a strong bimodal distribution corresponding to the fully occupied and empty sites. Importantly, such bimodal distributed forces with two strong peaks are very difficult to model even with neural network. On the other hand, as the electronic forces are nearly balanced by the elastic forces for  systems not too far from equilibrium, the total force $F_i$ is well characterized by a Gaussian-like distribution, which can be easily modeled by ML. 

Here we propose a ML model, summarized in Fig.~\ref{fig:NN}, for the direct prediction of the local total force $F_i$. As discussed previously, the locality principle implies that the effective force $F_i$ acting on the local lattice mode~$Q_i$ only depends on the structural configurations in the immediate neighborhood. 
Explicitly, the neighborhood centered at site-$i$ is defined as the collection of lattice distortions within a cutoff radius $r_c$:
\begin{eqnarray}
	\mathcal{C}_i = \left\{ Q_j \, \big| \, |\mathbf r_j - \mathbf r_i | \le r_c \right\}.
\end{eqnarray}
In general, the local environment is characterized by a set of feature variables $\bm G = \{G_1, G_2, \cdots \}$ which are obtained from the structural environment. For a given set of parameters for the electron Hamiltonian, the local force depends on the neighborhood $\mathcal{C}_i$ through a universal function of these feature variables:
\begin{eqnarray}
	F_i = \mathcal{F}\!\left( \bm G \right)= \mathcal{F}\bigl( \left\{ G_1(\mathcal{C}_i), G_2(\mathcal{C}_i), \cdots \right\}\bigr).
\end{eqnarray}
Importantly, here the complex dependence on the neighborhood is to be approximated by an ML model, which can be trained from exact solutions on small systems. It is worth noting that the ML approach described above essentially is to produce an effective classical force-field model in the adiabatic limit. Yet recent advances in ML, especially with deep-learning neural networks, offer a systematic way to derive this complex function $\mathcal{F}(\cdot)$ accurately and efficiently.

As shown in Fig.~\ref{fig:NN}, there are two central components of our proposed ML model:  a descriptor and a deep-learning neural network (NN). While the  universal expressive power of NN is utilized to achieve accurate approximation of the force field, the descriptor is introduced to preserve the symmetry of the original electron Hamiltonian. This is because despite the powerful approximation capability of NNs, symmetries of the original electron Hamiltonian can be learnt only statistically, but not exactly. A descriptor is introduced to provide a proper representation of the neighborhood configuration in such a way that the representation is itself invariant under transformations of the relevant symmetry group. By using these symmetrized feature variables as input to the NN, the ML prediction of the force or energy is guaranteed to be invariant with respect to the symmetry of the underlying electron models.

Descriptors also play a crucial role in ML-based quantum MD methods. Most molecular systems are invariant under translation and rotation operations as well as permutations of atomic species, representations of atomic neighborhood are expected to be invariant under these symmetry operations.  Numerous atomic descriptors have been proposed over the past decades to incorporate these basic symmetry properties into ML interatomic potential models~\cite{behler11,bartok13,ghiringhelli15,himanen20,rupp12,shapeev16,drautz19,hansen15,faber15,huo18}. A popular atomic descriptor used in many ML models is the atom-centered symmetry functions (ACSFs) built from relative distances and angles of atomic positions in the neighborhood~\cite{Behler2007,behler11}. A more systematic approach to build invariant feature variables is based on the so-called bispectrum coefficients, which are special triple-products of irreducible representations of the symmetry group~\cite{Bartok2010,bartok13}.

For condensed-matter systems, most of which are defined on specific lattices, the SO(3) rotational symmetry of free-space is reduced to discrete point-group symmetries. On the other hand, the dynamical degrees of freedom, such as local magnetic moments or order-parameters, are characterized by additional internal symmetry group. A general theory of descriptors for ML force-field models in condensed matter systems is recently discussed in Ref.~\cite{zhang22a}. In particular, a descriptor based on group-theoretical approach is presented; several explicit implementations have also been demonstrated in well-studied itinerant magnets and correlated electron systems~\cite{ma19,liu22,zhang22b,zhang20,zhang21a,zhang21b}. For application to the Holstein model, there is no internal symmetry associated with scalar lattice modes $Q_i$. Yet, the ML force field model needs to be invariant with respect to the $D_4$ point-group symmetry. Essentially, this means that the eight different lattice configurations that are related by symmetry operations of~$D_4$, as shown in the example of Fig.~\ref{fig:NN}, are to be mapped to  exact same feature variables~$\bm G = \{G_\ell\}$. 

To derive these symmetry-invariant feature variables, first we note that the neighborhood lattice configuration $\mathcal{C}_i$ forms a high-dimensional reducible representation of the $D_4$ group. It can then be decomposed into fundamental irreducible representations (IR's) of the point group. This decomposition can be highly simplified as the original representation matrix is automatically block-diagonalized, with each block corresponding to a fixed distance from the center-site. We use $ {\bm  f}^\Gamma = (f^\Gamma_1, f^\Gamma_2, \cdots , f^\Gamma_{D_\Gamma})$ to denote the basis function of IR of the symmetry-type $\Gamma$. For example, the lattice distortions of the four nearest-neighbor sites in Fig.~\ref{fig:NN} can be decomposed as $4 = 1A_1 + 1B_1 + 1E$, where $f^{A_1} = Q_a + Q_b + Q_c + Q_d$ and so on; more details can be found in Appendix~\ref{sec:descriptor}. Given these IR coefficients, one immediate class of invariants is their amplitudes $p^\Gamma = |{\bm f}^\Gamma|^2$, which is called the power spectrum of the representation. However, the descriptor also needs to account for crucial information on the relative phases of different IRs. 

A more general set of invariants which include relative phase information is the so-called bispectrum coefficients~\cite{Kondor}. These are special triple-products of IR coefficients with Clebsch-Gordan coefficients introduced to compensate the different transformation properties of the IRs. The collection of all bispectrum coefficients in principle provide a complete invariant description of the environment, which means they can be used to faithfully reconstruct the neighborhood configuration up to an arbitrary symmetry transformation. However, the rather large number of bispectrum coefficients implies a large set of feature variables, and often with huge redundancy. To circumvent this issue, we introduce the concept of reference IR coefficients ${\bm f}^{\Gamma}_{\rm ref}$, which are obtained by applying similar decomposition procedure to large symmetry-related blocks of $\mathcal{C}_i$ such that they are insensitive to small variations of the neighborhood~\cite{zhang22a}. Importantly, the relative phase of two IRs can be restored from their respective relative phases to the reference IR, e.g. $\eta^{\Gamma} \sim {\bm f}^{\Gamma} \cdot {\bm f}^{\Gamma}_{\rm ref}$.  A complete set of invariant feature variables is then given by the power spectrum $p^\Gamma$ and the phases $\eta^\Gamma$; see Appendix~\ref{sec:descriptor} for more details.

\begin{figure}
\centering
\includegraphics[width=0.99\columnwidth]{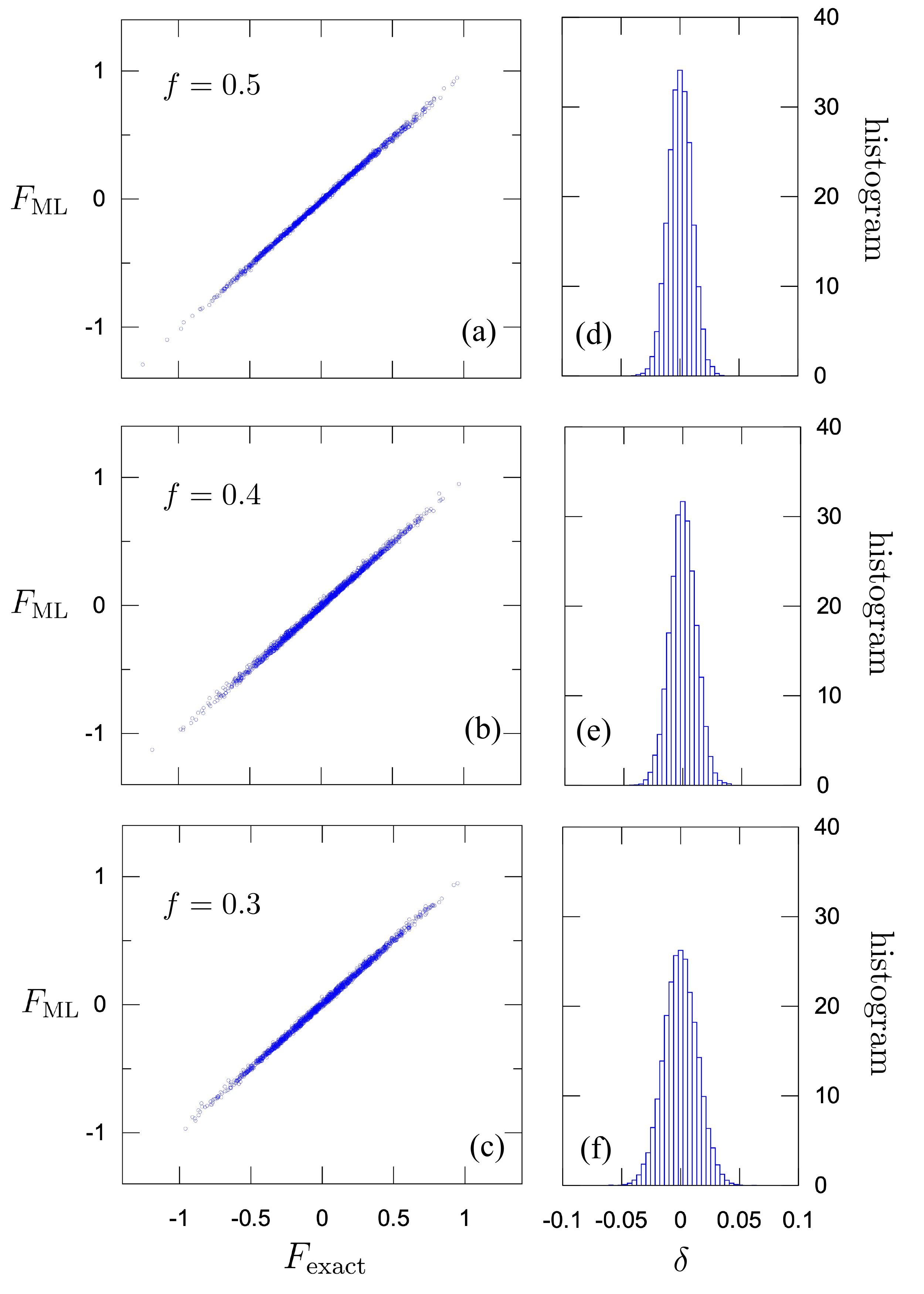}
\caption{Benchmark of ML force-field model for adiabatic dynamics of Holstein model. Panels~(a)--(c) on the left show the ML predicted force $F_{\rm ML}$ versus the exact force $F_{\rm exact}$ for three different electron filling fractions $f = 0.5$, 0.4 and 0.3. The corresponding histograms of the prediction error $\delta = F_{\rm ML} - F_{\rm exact}$ are shown on panels~(d)--(f) on the right. The standard deviations for the three filling fractions are $\sigma =$ 0.010, 0.011 and 0.014 (from top to bottom).}
\label{fig:force_benchmark}
\end{figure}

\begin{figure}
\centering
\includegraphics[width=0.99\columnwidth]{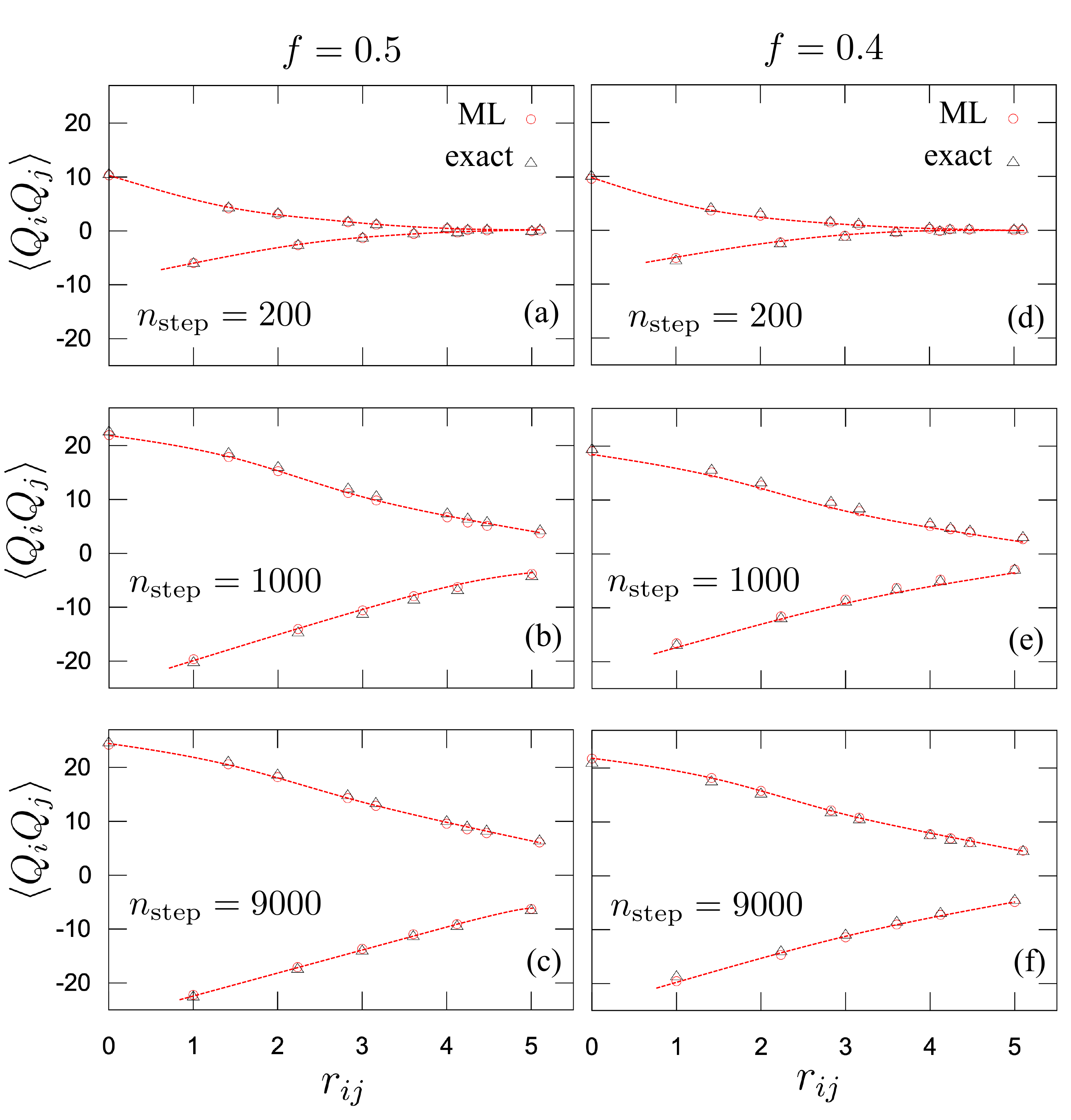}
\caption{Comparison of lattice correlation functions $C_{ij} = \langle Q_i Q_j \rangle$ obtained from Langevin simulations with the ML force-field model and the ED method. Thermal quench simulations of a $40\times 40$ system at two different filling fractions $f = 0.5$ (left) and $f = 0.4$ (right) were carried out to produce these correlation functions at various time steps after the quench. }
\label{fig:corr}
\end{figure}

The resultant invariant representation of $\mathcal{C}_i$ is then fed into the NN which in turn produces the scalar force $F_i$ at its output node. A five-layer NN model is constructed and trained using PyTorch~\cite{Paszke2019, Nair2010, Barron, Paszke2017, He2015, Kingma}. The training datasets are obtained from the ED solutions of a mixture of random configurations and quasi-ordered CDW states on a small $40\times 40$ system. To properly capture configurations during the relaxation process, intermediate states obtained from thermal quench simulations are included in the training dataset. More details on the NN structure and training process are discussed in Appendix~\ref{sec:NN}.

We have built three ML models corresponding to different electron filling fractions $f = 0.5$, 0.4 and 0.3 of the Holstein model. Assuming dimensionless lattice variables~$Q_i$, the model parameters of the electron Hamiltonian all have unit of energy. By setting $t_{\rm nn} = 1$, which serves as the energy unit, the other parameters of the Holstein model are $g = 3.5$, $k=1$, and $\kappa=0.18$. The lattice dynamics is characterized by the fundamental frequency $\omega = \sqrt{k / m}$ of the simple harmonic oscillator; its inverse $\omega^{-1}$ can thus be used as the time unit. The dissipation timescale is given by $\tau = \gamma / k$, and we have used a damping coefficient $\gamma$ such that the dimensionless dissipation in one fundamental cycle is $\omega \tau \sim 0.09$. Finally, a time-step $\Delta t = 0.022\, \omega^{-1}$ is used in all Langevin dynamics simulations discussed in this work.

For each model, 600 configurations of~$Q_i$ on the $40 \times 40$ square lattice are used in the training dataset. Specifically, the neighborhood configuration $\mathcal{C}_i$ and the corresponding force $F_i$ from ED of each lattice site constitutes one training data in this supervised learning. This means that the total number of dataset is $1600\times 600 = 960,000$. Remarkably, even with this moderate size of training dataset, rather accurate force predictions were achieved for all three models, as shown in Fig.~\ref{fig:force_benchmark}. The distribution of the prediction error $\delta = F_{\rm ML} - F_{\rm exact}$ is shown on the right panels, where a rather small mean-square error is obtained for all three cases.

Next we incorporated the trained ML models into the Langevin dynamics method and performed thermal-quench simulations of the Holstein model. The results were then compared with ED-Langevin simulations to benchmark whether the ML models can also reproduce the dynamical evolution of the Holstein model.  An initially random state is suddenly quenched to a temperature $T = 0.1$ at time $t = 0$. We computed the correlation function between two local lattice amplitudes $C_{ij} = \langle Q_i Q_j \rangle$ at various times after the thermal quench. Fig.~\ref{fig:corr} shows the comparison of the correlation functions obtained from ML and ED Langevin simulations for two different electron filling fractions. To circumvent statistical fluctuations due to small lattice sizes, each correlation function was computed by averaging over 30 independent runs. 

The correlation functions exhibit a short-period oscillation which is enveloped by two gradually decaying curves. The oscillation is due to the staggered lattice distortions $Q_i \sim (-1)^{x_i + y_i}$ that accompany the checkerboard charge modulation. A correlation length can be estimated from the two envelop functions. Interestingly, short-range CDW correlation emerges rather quickly, e.g. at $n_{\rm step} = 200$, after a thermal quench to $T = 0.1$. Yet, comparison of the correlation functions at large time steps shows that the build-up of longer-range checkerboard order is rather slow at this temperature. As will be discussed in the next section, this is related to a power-law domain growth with a relatively small exponent. Importantly, as shown in Fig.~\ref{fig:corr}, excellent agreement between ML and ED simulations was obtained at different times of the  relaxation process. This provides strong evidence that, in addition to accurate force predictions, our ML models also faithfully capture the dynamics of the Holstein model.

While our ML model is designed to predict the total force $F_i$, it also serves as a model for computing the local electron density $n_i = \langle c^\dagger_i c^{\,}_i \rangle$ from the structural environment~$\mathcal{C}_i$. This is because the electronic force $F_{\rm elec}$ is linearly proportional to $n_i$ in the Holstein model as shown in Eq.~(\ref{eq:F_elec}). And since the elastic contribution can be trivially computed from on-site and nearest-neighbor~$Q_j$ using Eq.~(\ref{eq:F_elastic}), by subtracting it from the total force, one can obtain an accurate estimate of the electron force, hence the local electron density.  We note in passing that, as a predictor of $n_i$, our ML model is transferrable in the sense that it can be used in different Holstein models with same electron Hamiltonian, yet different classical parameters.

Finally, as a further comparison, we have also implemented a BP-type ML model where the output of the NN is the local energy $\epsilon_i$ associated with individual lattice sites. The local force is then computed from the total energy through automatic differentiation of the NN. We found that, with similar descriptor, cutoff radius, hyperparameters of the NN, and training dataset, the ML model with direct force prediction gives a much better accuracy than the energy-based BP-type model.  This difference of the two approaches might be attributed to a better locality of the effective force $F_i$ than that of the site-energy $\epsilon_i$. For example, force is directly observable, hence is uniquely defined in the model. On the other hand, more complex NN structure or training dataset might be required in order to capture the implicit partitioning of the total energy into local contributions.

\section{Machine learning dynamical simulation of CDW coarsening} 

\label{sec:result}

\begin{figure*}
\centering
\includegraphics[width=1.99\columnwidth]{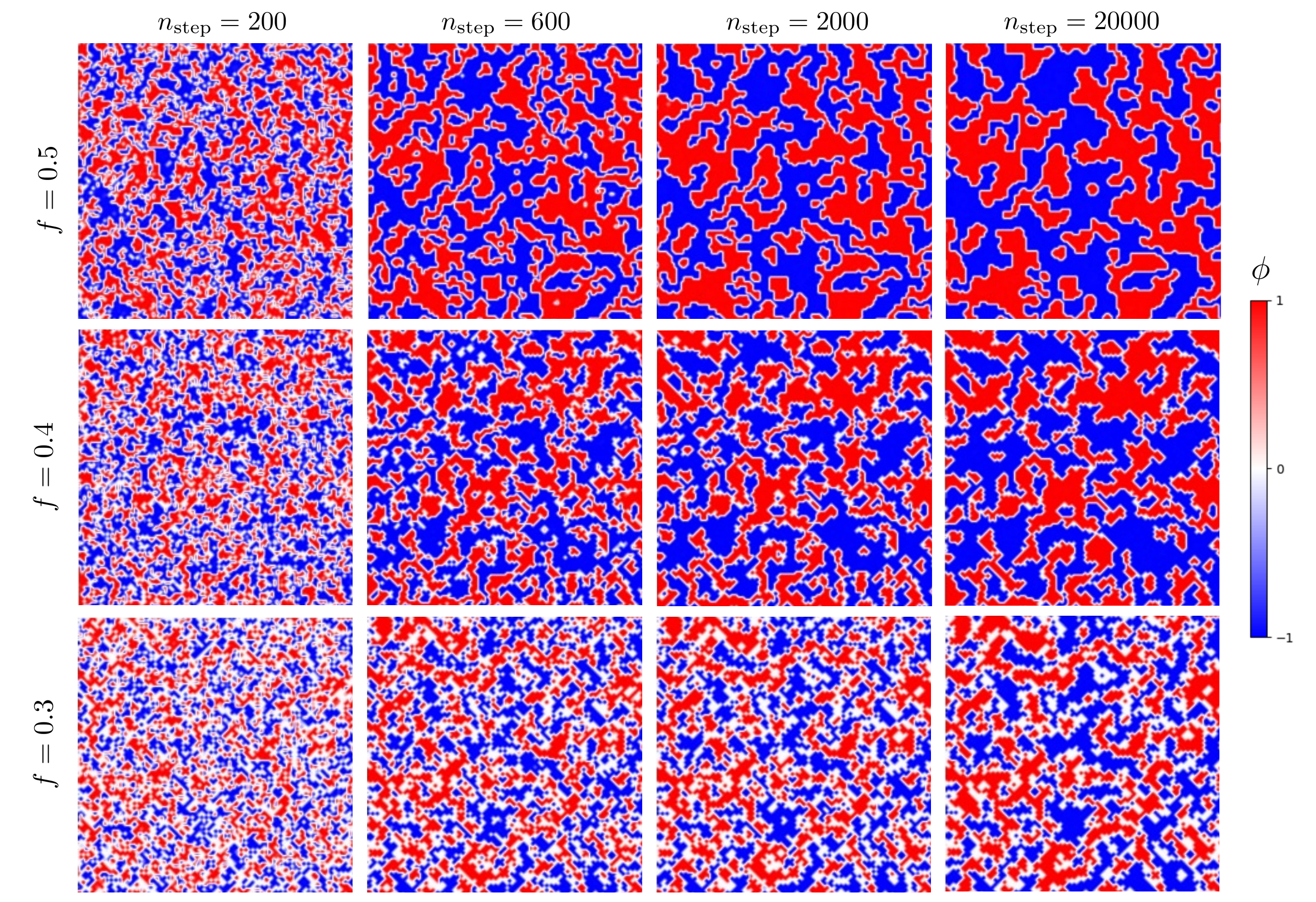}
\caption{Snapshots of local CDW parameter $\phi_i$ at various time steps after a thermal quench of the Holstein model. An initially random configuration is suddenly quenched to a temperature $T = 0.1$ at time $t = 0$ ($n_{\rm step} = 0$). The ML-Langevin dynamics was used to simulate the relaxation of the system toward equilibrium. The red and blue regions correspond to CDW domains with order parameter $\phi = +1$ and $-1$, respectively. A time step $\Delta t = 0.022 \, \omega^{-1}$ is used in the simulations, where $\omega = \sqrt{k/m}$ is the fundamental frequency of the lattice degrees of freedom.   }
\label{fig:config}
\end{figure*}

As discussed in Sec.~\ref{sec:intro}, the high efficiency and linear scalability are features of ML models that make large-scale dynamical simulations possible. As a demonstration, Langevin dynamics simulation based on ED for the force calculation of a $40\times 40$ Holstein model took about 16 hours for $10,000$ steps. On the other hand, Langevin simulations of exact the same size and time-steps only took $\sim $1~minute using the ML force field model. This corresponds to a $1000$-fold improvement in efficiency. Extrapolating to simulations on a $200\times 200$ lattice to be discussed below based on the $\mathcal{O}(N^3)$ scaling, the ML-based Langevin simulation is expected to be roughly $10^6$ times faster than that based on the ED method.

Here we apply the ML-Langevin dynamics to study the phase ordering of CDW order of the Holstein model. To this end, we performed the thermal quench simulations where an initially state with random local distortions was suddenly cooled to a low temperature $T = 0.1$ at time $t = 0$. The Langevin simulations were carried out on a $200\times 200$ system with the same model parameters as those used to generate the training dataset discussed in the previous section. Three different filling fractions $f = 0.5$, 0.4, and 0.3 are considered in order to investigate the effects of hole doping on the domain structures and growth laws. As discussed above, the low-temperature CDW phase of the Holstein model is characterized by a broken $Z_2$, or sublattice, symmetry. The checkerboard charge modulation of a perfect CDW order can be described by a parameter $\delta$:
\begin{eqnarray}
	\label{eq:n_mod}
	n_i = \frac{1}{2} \left[1 + \delta \, \exp\left({i \mathbf Q \cdot \mathbf r_i}\right) \right]
\end{eqnarray}
where $\mathbf Q = (\pi, \pi)$ is the ordering wave vector, and the phase factor $\exp({i \mathbf Q \cdot \mathbf r_i}) = \pm 1$ for the $A$ and $B$ sublattices, respectively.  The $Z_2$ symmetry transform a CDW order with parameter $+\delta$ to one with $-\delta$.  As the system relaxes toward the equilibrium state after the quench, multiple CDW domains of opposite signs of charge-modulation develop simultaneously. The dynamics of phase ordering is thus dominated by the merging and growth of these CDW domains. To characterize the inhomogeneous intermediate states during the relaxation process, we introduce a local CDW order parameter
\begin{align}
	\phi_i = \Bigl(n_i - \frac{1}{4}\sum_j\phantom{}^{'} n_j \Bigr) \exp\left({i \mathbf Q \cdot \mathbf r_i}\right), 
\end{align}
where the prime in the second term indicates that the summation is restricted to the nearest neighbors of site-$i$. This local parameter essentially measures the difference of the electron number at a given site and that of its nearest neighbors. A nonzero $\phi_i$ thus indicates the presence of local charge modulation around site-$i$. The long-range charge modulation described in Eq.~(\ref{eq:n_mod}) corresponds to a uniform order parameter $\phi_i = \delta$.

Fig.~\ref{fig:config} shows snapshots of the local CDW order $\phi_i$ at various time steps after the thermal quench. As discussed in Sec.~\ref{sec:ML}, the ML model is used to also compute the local electron density $n_i$ form a given snapshot of lattice distortions. The red and blue regions, corresponding to $\phi_i =+1$ and $-1$, respectively, are CDW domains related by the $Z_2$ symmetry. The two types of CDW domains are separated by interfaces of vanishing $\phi_i$, corresponding to the white regions.  The emergence of numerous red and blue domains at small time steps indicates that CDW order with strong charge modulation is quickly established after the quench. Yet the correlation length of the CDW order is rather short at the early stage of phase ordering, as evidenced by the relative small sizes of these CDW domains.  As the system relaxes toward equilibrium, these CDW domains merge into bigger ones, giving rise to a coarser mixture of the two ordered phases.

\begin{figure}
\centering
\includegraphics[width=0.99\columnwidth]{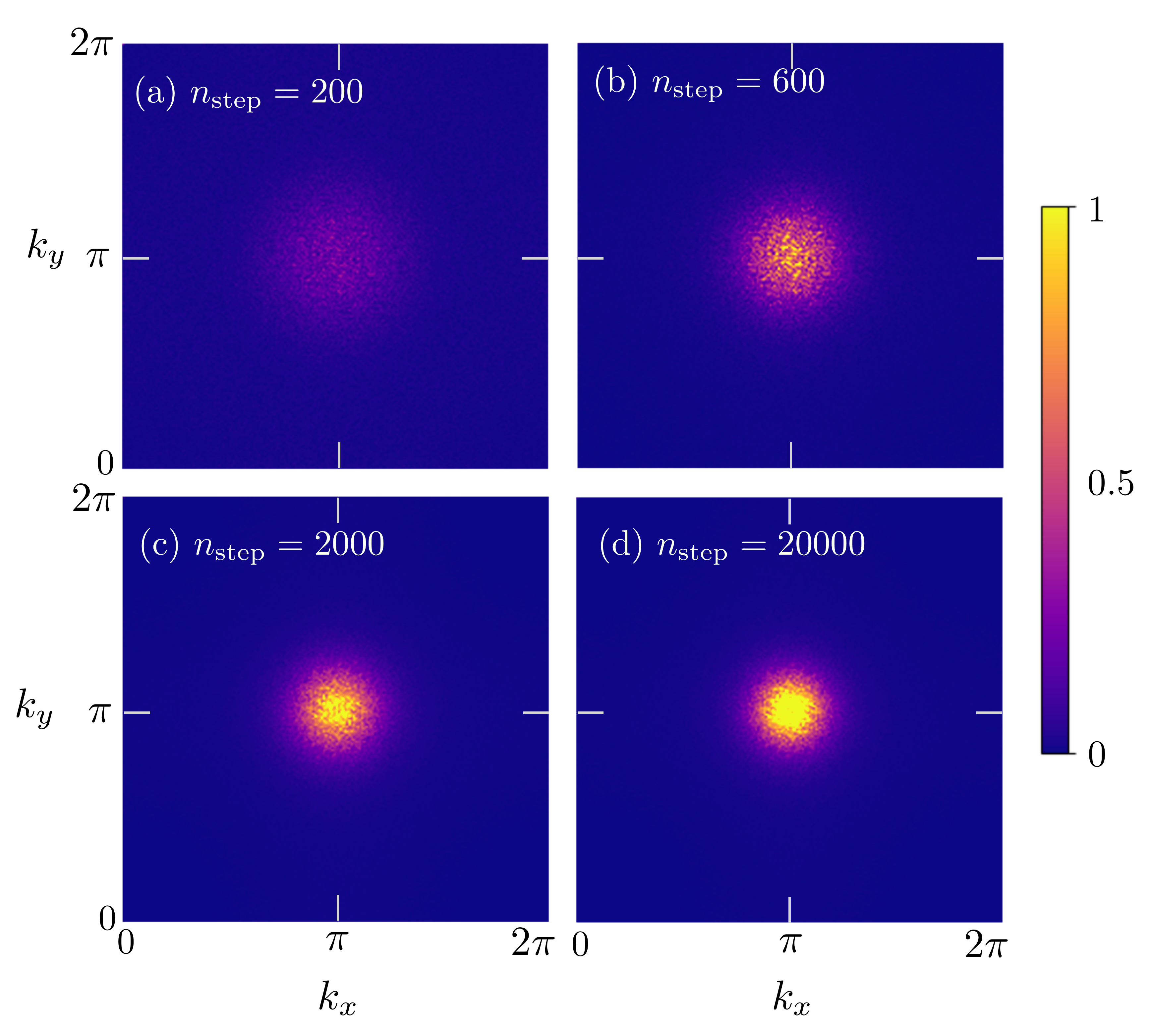}
\caption{Structure factor $S(\mathbf k, t)$ of the electron density distribution $n(\mathbf r_i, t) = n_i(t)$ at various time steps after a thermal quench from random configurations to a temperature $T = 0.1$. The system size is $200\times 200$ and the filling fraction is $f = 0.5$. }
\label{fig:sk}
\end{figure}

\begin{figure}
\centering
\includegraphics[width=0.99\columnwidth]{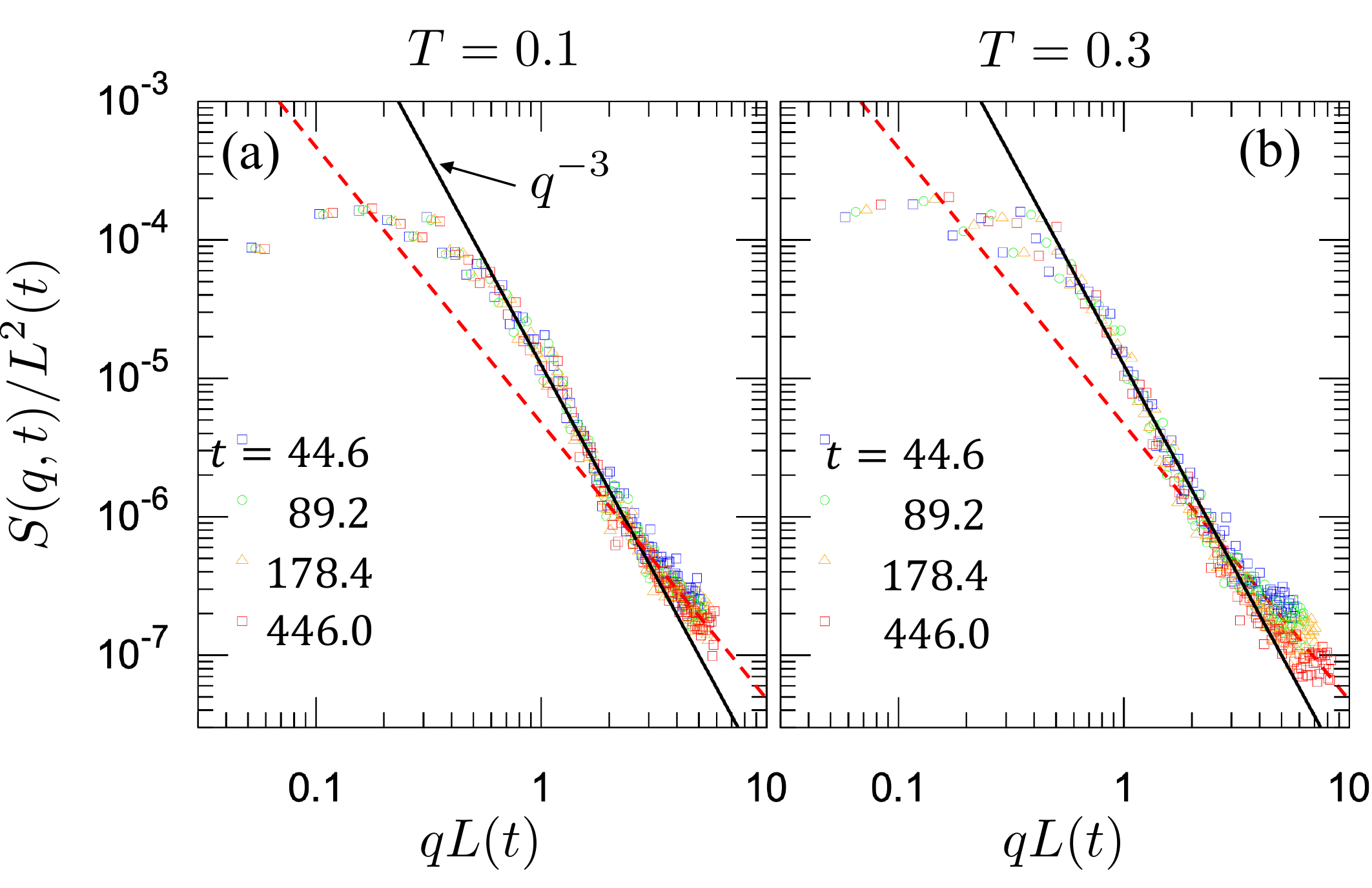}
\caption{Rescaled structure factor $S(q, t) / L^2(t)$ versus the dimensionless wave vector $q L(t)$ at different times for two different quench temperatures: (a) $T = 0.1$ and (b) $T = 0.3$. The filling fraction is set at $f = 0.5$. The black line shows the $q^{-3}$ tail of the 2D Porod’s law. The red dashed line marks the power law $q^{-\gamma}$, with an exponent $\gamma \sim 2$.}
\label{fig:Sk_T}
\end{figure}

To quantify the coarsening of CDW domains, we first compute the structure factor of the CDW state:
\begin{eqnarray}
	S(\mathbf k, t) = \left| \tilde{n}(\mathbf k, t) \right|^2,
\end{eqnarray}
where $\tilde{n}(\mathbf k, t)$ is the Fourier transform of the time-dependent electron density $n(\mathbf r_i, t) = n_i(t)$:
\begin{eqnarray}
	\tilde{n}(\mathbf k, t) = \frac{1}{N} \sum_i \left( n(\mathbf r_i, t) - \bar{n} \, \right) \exp\left({i \mathbf k \cdot \mathbf r_i} \right),
\end{eqnarray}
Here $\bar{n}$ is the average number of electrons per site, which is the same as the filling fraction $f$. The structure factors at various time steps after a quench to $T = 0.1$ are shown in Fig.~\ref{fig:sk}. The emergence of checkerboard charge patterns corresponds to a peak at the wave vector $\mathbf Q=(\pi, \pi)$. However, contrary to Bragg peaks that are characteristic of long-range order, the quenched states here exhibit a broad diffusive peak due to the coexistence of multiple CDW domains of opposite signs.  As the system equilibrates, the coarsening of CDW domains results in a stronger and sharper peak at $\mathbf Q$. The width of the diffusion peak thus provides a quantitative estimate of the average size $L(t)$ of the CDW domains at time $t$. Specifically, it is defined as
\begin{eqnarray}
	L^{-1}(t) = \sum_{\mathbf k} S(\mathbf k, t) \left| \mathbf k - \mathbf Q \right| \Big/ \sum_{\mathbf k} S(\mathbf k, t).
\end{eqnarray}
In addition to a measure of typical domain sizes, $L(t)$ can also be viewed as the correlation length of the CDW states. By properly rescaling the time-dependent structure factor and the wave vector using this characteristic length, the data points at different times collapse in the vicinity of a hidden curve, as shown in Fig.~\ref{fig:Sk_T} for two different quench temperatures. This indicates that the coarsening of CDW domains exhibits a dynamical scaling 
\begin{eqnarray}
	\label{eq:sqt}
	S(q, t) L^2(t) = \mathcal{G}\bigl( q L(t) \bigr),
\end{eqnarray}
where $q = |\mathbf k - \mathbf Q|$ is the distance from the CDW peak in momentum space, and $\mathcal{G}(x)$ denotes the hidden universal scaling function.  Dynamical scaling has been observed in the phase ordering of numerous Ising-type transitions. In many cases, the scaling function exhibits a $1/q^{d+1}$ power-law behavior at large wave vectors, where $d$ is the spatial dimension. This universal power-law dependence, also known as the Porod's law~\cite{Bray1994, Puri2009}, can be attributed to the rather sharp interfaces that separate the two ordered states related by the $Z_2$ symmetry. For our case of 2D Holstein model, the structure factor seems to be well described by the $q^{-3}$ power law for intermediate values of~$q$; see Fig.~\ref{fig:Sk_T}. However, significant deviation from the Porod's law can be seen for large values of the wave vector, as shown by the red dashed lines in Fig.~\ref{fig:Sk_T}, which denote a power-law $q^{-\gamma}$ with an exponent $\gamma \sim 2$. As will be discussed below, this deviation is related to the complex domain-wall structures of the Holstein model.

The dynamical scaling Eq.~(\ref{eq:sqt}) also means that the coarsening of CDW order is characterized by a single characteristic length $L(t)$. The kinetics of phase ordering can thus be characterized by the time dependence of this length scale. As the CDW transition belongs to the Ising universality class, the coarsening of CDW domains is expected to be similar to that of standard Ising systems. Kinetic Monte Carlo simulations of the nearest-neighbor ferromagnetic Ising model on various lattices find a power-law domain growth~\cite{Bray1994, Puri2009}
\begin{align}
	L(t) \sim t^{\alpha},
	\label{eq:Lt_power}
\end{align}
where the growth exponent $\alpha$ is a universal value independent of lattice geometries and dimensionality. The exponent, on the other hand, depends on whether the dynamics conserve the Ising order parameter. For non-conserved dynamics as in our case of CDW ordering, the universal exponent is $\alpha = 1/2$, and the corresponding power-law growth is also known as the Allen-Cahn law~\cite{Bray1994, Puri2009}. More generally, the same power-law behavior can also be obtained from the relaxational model-A dynamics of the coarse-grained Ising order parameter $\phi(\mathbf r)$, which is described by the time-dependent Ginzburg-Landau (TDGL) equation. Analytical calculation of TDGL assuming a random initial state leads to an equal-time correlation function of the scaling form $\langle \phi(\mathbf 0, t) \phi(\mathbf r, t) \rangle \sim \mathcal{K}(|\mathbf r| / L(t))$~\cite{ohta82}, where $\mathcal{K}(x)$ is a universal scaling function and the correlation length $L(T)$ follows the Allen-Cahn power law.

\begin{figure}
\centering
\includegraphics[width=0.99\columnwidth]{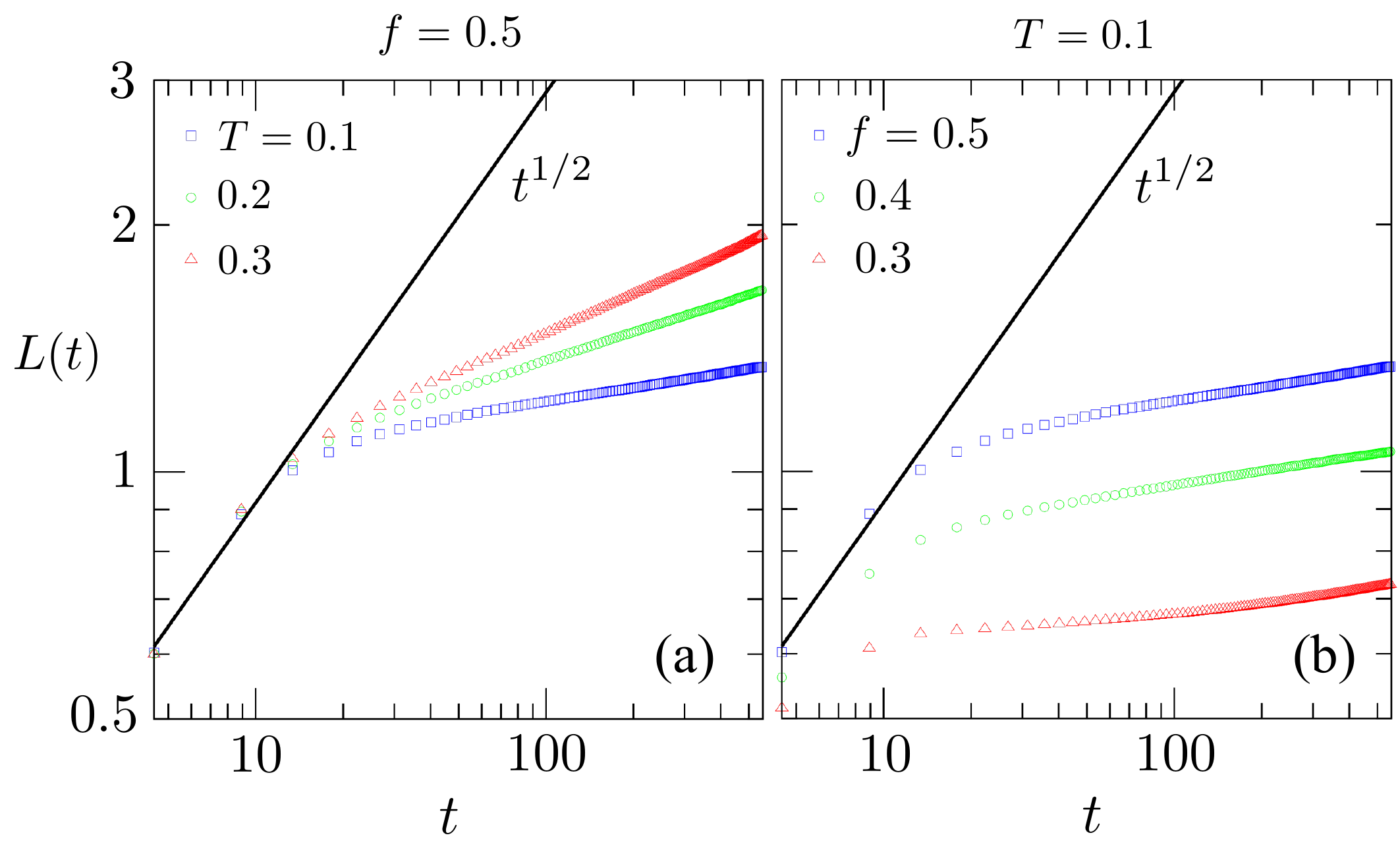}
\caption{The characteristic length $L(t)$ of CDW order as a function of time obtained from the ML-Langevin simulations. Panel (a) shows the $L(t)$ curves of three different temperatures at the filling fraction $f = 0.5$. The power-law growth at late stage is characterized by an exponent $\alpha = 0.059$, 0.115, and 0.155 for quench temperatures $T = 0.1$, 0.2, and 0.3, respectively. (b) $L(t)$ curves at filling fractions $f = 0.5$, 0.4 and 0.3 for thermal quench to $T = 0.1$. The late-stage power-law growth of all three electron fillings can be well approximated by an exponent $\alpha = 0.059$. Also shown for reference is the Allen-Cahn growth law with exponent $\alpha = 1/2$.}
\label{fig:Lt}
\end{figure}

Coming back to the coarsening of CDW order, the time dependence of the characteristic length $L(t)$ computed from ML-Langevin simulations is shown in Fig.~\ref{fig:Lt}(a) for three different quench temperatures at the filling fraction $f = 0.5$. Interestingly, while the growth of the CDW domains at late stage indeed exhibits a power-law behavior, the dynamical exponent $\alpha$ is non-universal and temperature dependent. Moreover, the extracted exponents at the three simulated temperatures, $\alpha_{T = 0.1} = 0.059$, $\alpha_{T = 0.2} = 0.115$, and $\alpha_{T = 0.3} = 0.155$, are significantly smaller than the Allen-Cahn exponent $1/2$, which is expected for a non-conserved Ising order parameter. This indicates a much slower phase ordering than the conventional Ising transition. On the other hand, for a given quench temperature, the dynamical exponent $\alpha$ is almost independent of the electron filling fraction $f$, as shown in Fig.~\ref{fig:Lt}. However, the growth rate, i.e. the prefactor of the power-law time dependence, is reduced as the system is doped away from half-filling.

\begin{figure}
\centering
\includegraphics[width=0.95\columnwidth]{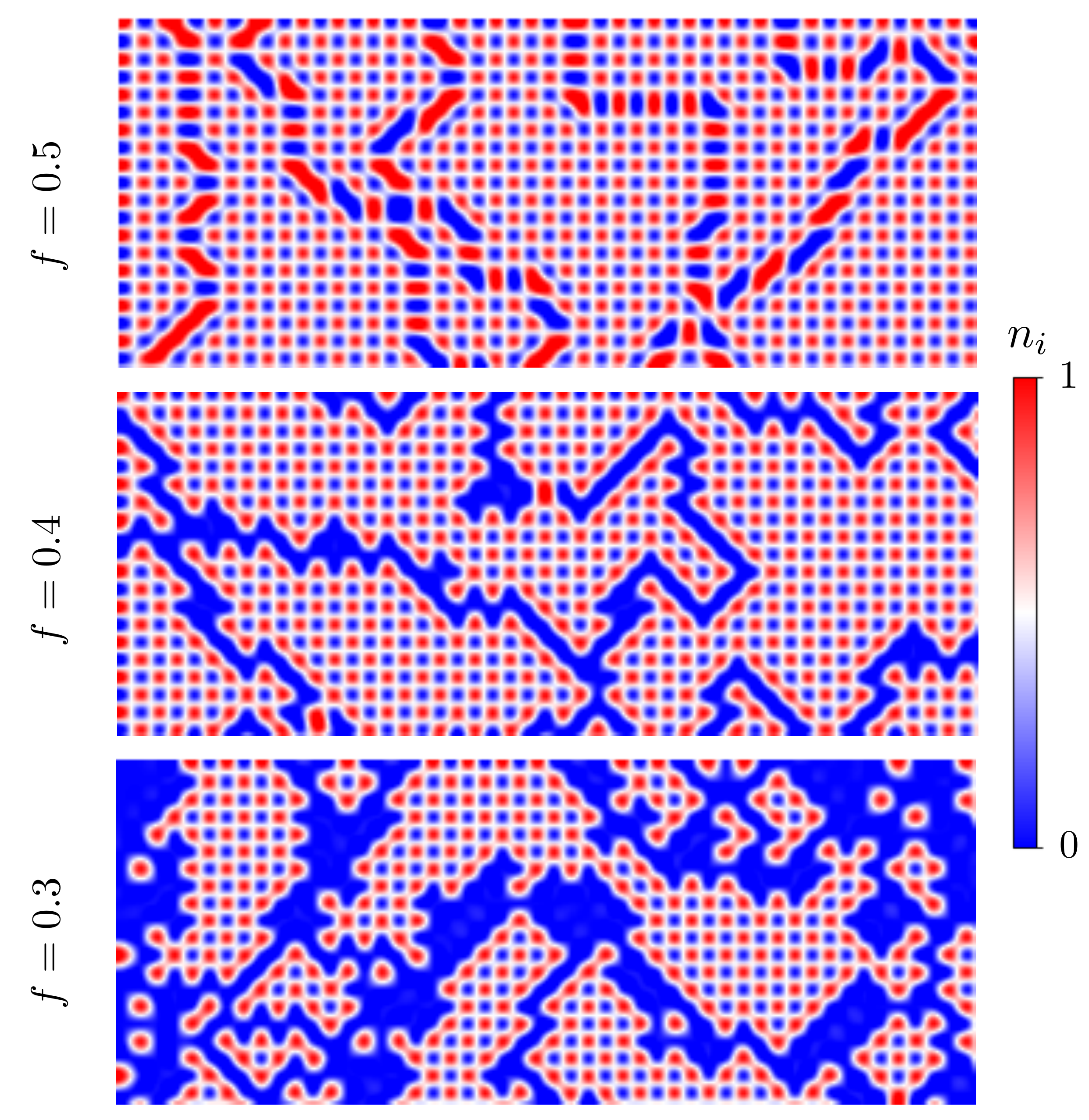}
\caption{Snapshots of local electron density $n_i$ at the late-stage ($n_{\rm step} = 20,000$) of CDW coarsening. The filling fractions from top to bottom are $f = 0.5$, 0.4, and 0.3.}
\label{fig:dw}
\end{figure}

To understand the unusual power-law behavior of CDW coarsening, we note that the expansion of an ordered region is essentially controlled by the structure and dynamics of the interface, or domain wall, that separates the two domains related by the $Z_2$ symmetry. Indeed, the Allen-Cahn power-law can be understood from a domain-wall motion whose velocity is proportional to the curvature of the interface.  Approximating the velocity by the domain growth rate $v \sim dL/dt$, and the curvature by the inverse domain size $\kappa \sim 1/L$, we obtain an equation of motion $dL/dt \sim -1/L$, from which one can easily derive a power-law growth with exponent $\alpha = 1/2$. A more rigorous derivation based on TDGL gives the same result. 

The interface dynamics in the TDGL theory, or in standard nearest-neighbor Ising models, is entirely determined by the conformation of the order-parameter field. In general, the domain-walls are rather sharp with relatively simple structures, especially at late stages of phase ordering. On the other hand, due to the involvement of the electron degrees of freedom, domain walls in the CDW states of Holstein model exhibit rather complex structures. This is demonstrated in Fig.~\ref{fig:dw} which shows a close-up view of local electron density of late-stage CDW states at three different filling fractions. For example, in the half-filling case shown in the top panel of Fig.~\ref{fig:dw}, the domain walls consist of alternating segments of fully-occupied and empty sites. Consequently, domain-wall motion necessarily requires rearrangements of these segments, giving rise to complex dynamics that is beyond the Allen-Cahn theory. 

For systems with a reduced filling fraction, the doped holes are expelled from the CDW domains similar to the scenario of doped Mott insulators. Consequently, the phase ordering of  hole-doped Holstein models can also be viewed as a phase separation process. It is worth noting that the hole-rich phase here is mostly confined to the interface region that separates CDW domains of opposite signs. This is in stark contrast to phase-separated states in correlated electron systems, such as the double-exchange or Hubbard models, where the doped holes aggregate in small clusters which then merge with each other as the system equilibrates. Importantly, the expansion of CDW domains is a complex process that involves the deformation, breaking, and reconnection of the hole-rich interface. Also contrary to the isotropic interfaces in the TDGL theory, the hole-rich domain walls of the CDW states in Fig.~\ref{fig:dw} clearly favor the diagonal directions of the square lattice.

\section{Conclusion and outlook} 

\label{sec:summation}

In summary, we have presented a scalable ML force-field model for the adiabatic CDW dynamics in electron-lattice coupled systems. Within the adiabatic approximation,  the dynamical evolution of a CDW state is governed by the lattice dynamics with a driving force computed from a quasi-equilibrium electron liquid. Assuming locality principle for the electronic forces, a multi-layer NN model is trained to accurately approximate the complex dependence of the force on local lattice configurations. Additionally, a lattice descriptor is developed to incorporate the symmetry of the electron Hamiltonian into the ML model.  We demonstrate our approach by applying it to study the phase-ordering dynamics in the semi-classical Holstein model on the square lattice. Our ML model trained by exact solution of $40\times 40$ systems not only accurately predict the driving forces, but also faithfully capture the dynamical evolution of the Holstein model. Also importantly, compared with the exact diagonalization for the force calculation, significant improvement of efficiency is expected for dynamical simulations with the trained ML model.

By incorporating the ML model into the Langevin dynamics method, we have performed large-scale simulations of the coarsening dynamics of CDW order in the semi-classical Holstein model.   While numerous studies have firmly established the Ising universality class for CDW transition in the Holstein model, very little is known about its phase-transition kinetics.  Intriguingly, our large-scale simulations uncovered a non-universal power-law domain growth that is contrary to the expected Allen-Chan law for non-conserved kinetics of the Ising order. Instead, the growth exponent is shown to be temperature dependent, and almost independent of the electron filling fractions. These unusual behaviors likely can be attributed to the complex structure of domain-walls of the CDW order and the nontrivial interplay of the lattice and electrons during domain coarsening. In fact, the temperature dependence of the growth exponent resembles that of the random-field Ising model~\cite{corberi12,corberi15},  despite the absence of quenched disorder in our simulations. This similarity suggests a self-generated dynamical disorder due to the interplay between electrons and CDW order. A careful study of this anomalous coarsening dynamics is left for the future. 



Thanks to the relative simplicity of  lattice variables in the Holstein model, our ML model is built to directly predict the local forces. Nonetheless, our proposed formalism can be directly generalized to energy-based ML models such as the Behler-Parrinello scheme. Comparison of the two approaches based on similar NN structure and training datasets showed that the direct-force ML model offers a significantly better accuracy to the energy-based model.   However, BP-type ML methods which focus on the prediction of a local energy provide a more general approach for CDW phases with complex lattice distortions, such as doublet Jahn-Teller distortions. On the other hand, since forces are computed from energy derivatives in the BP-type scheme, energy-based ML models are thus restricted to the modeling of conservative forces~\cite{PZhang_arx12124}. One crucial advantage of the direct-force ML model is the capability to describe non-conservative electronic forces arising from driven CDW systems~\cite{hollander15,vaskivskyi16,geremew19,zheng17,SZhang_arx02194}. As the model proposed in this work is only applicable to scalar force field of Holstein-type models, further development are required for the ML modeling of multi-component non-conservative forces.


Finally, CDW order in the Holstein model mainly arise from the electron-lattice coupling. As a result, the lattice dynamics plays the dominant role in the adiabatic evolution of CDW order in the Holstein model. For more complicated systems where charge modulation is at least partially stabilized by electron-electron interactions, the evolution of the CDW order parameter, which describe the collective electron behaviors, might be described by its own equation of motion. For example, coupling to a heat reservoir could leads to a dissipative model-A dynamics $\partial \phi_i/ \partial t = -\eta \partial \langle \mathcal{H}_{\rm CDW} \rangle / \partial \phi_i$, which is coupled to the lattice dynamics. Here $\mathcal{H}_{\rm CDW}$ describes an effective electron Hamiltonian, e.g. as in the Hartree-Fock approximation, with the introduction of the CDW order parameter. Consequently, in addition to the force on lattice variables, an effective force on the CDW order $F^{\rm CDW}_i = -\partial \langle \mathcal{H}_{\rm CDW} \rangle / \partial \phi_i$ is also required for a complete dynamical description. The ML framework presented here can be generalized to include the order-parameter dynamics. Essentially, the neighborhood configuration $\mathcal{C}_i$ now includes both lattice and local CDW order parameters, which are then used to predict both forces via a neural network.


\begin{acknowledgments}
This work was supported by the US Department of Energy Basic Energy Sciences under Award No. DE-SC0020330. The authors also acknowledge the support of Research Computing at the University of Virginia.
\end{acknowledgments}

\appendix

\section{Lattice descriptor}

\label{sec:descriptor}

As discussed in Sec.~\ref{sec:NN}, the goal of a lattice descriptor is to preserve lattice symmetry of the original lattice Hamiltonian in the ML model.  In the case of square lattice, the lattice descriptor maps the eight configurations related by the $D_4$ symmetry operations into a symmetry-invariant generalized coordinates $\{ G_\ell \}$. One systematic approach to obtain these invariants is based on the group-theoretical method~\cite{Mamermesh}.
To this end,  we first note that the local distortions $\{ Q_i \}$ in a given environment $\mathcal{C}_i$ form a high-dimensional representation of the $D_4$ point group. This neighborhood representation can then be decomposed into  irreducible representations (IRs)~\cite{Mamermesh}. This decomposition is considerably simplified due to the lattice geometry. Essentially, since the distance between a neighborhood site-$j$ and the center site-$i$ is invariant under operations of the $D_4$ group, the neighborhood representation $\mathcal{C}_i$ is already block-diagonalized, with each block corresponding to a layer of neighbors sharing the same distance to the center. 

\begin{figure}
\centering
\includegraphics[width=0.95\columnwidth]{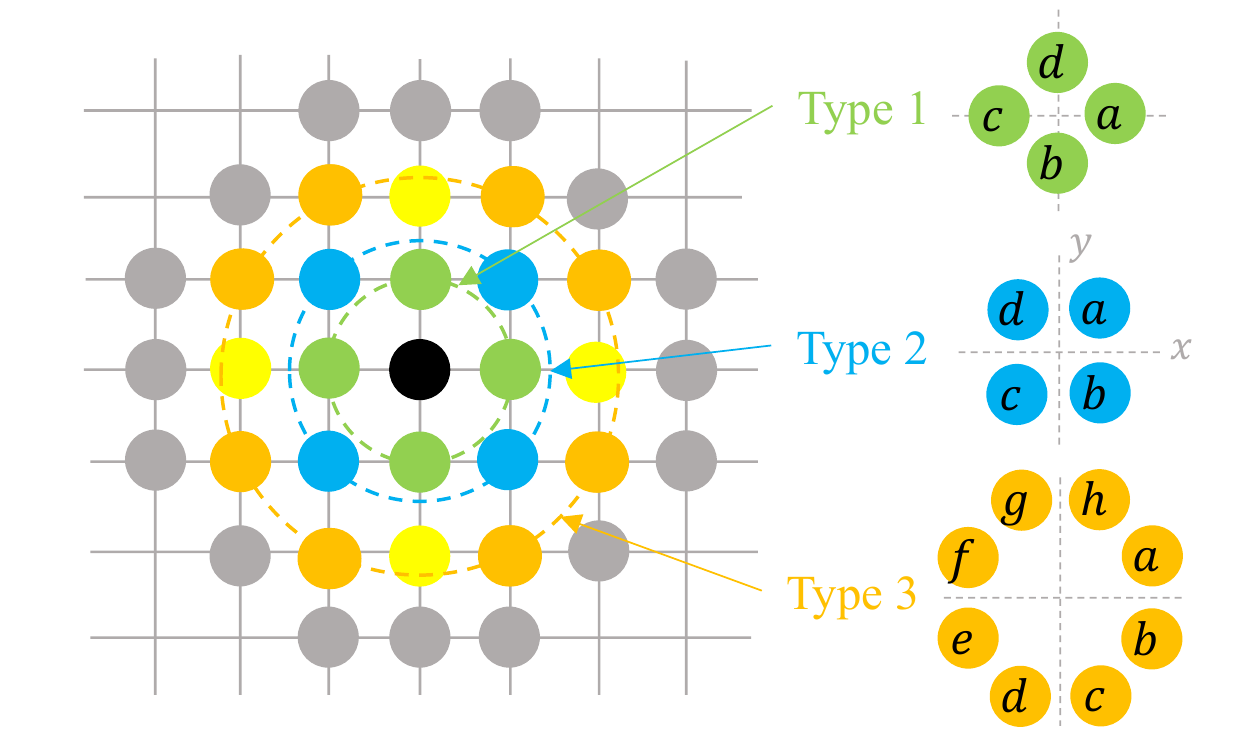}
\caption{Schematic diagram showing the invariant blocks of the neighborhood representation. Due to the lattice geometry, the neighbors of a given site at the center can be grouped into many layers by their distances to the center. Different layers are outlined by the dashed circles in the figure. The local distortions $\{ Q_i \}$ in the neighborhood $\mathcal{C}_i$ form a high-dimensional basis for the $D_4$ point group. The basis functions in each layer correspond to a reducible representation, there are three types that the representations are inequivalent.}
\label{fig:descriptor}
\end{figure}

For the square lattice, these invariant blocks can be classified into three types illustrated in Fig.~\ref{fig:descriptor}. The first two types are both 4-dimensional representations of $D_4$, and can be decomposed as $4 =A_1 \bigoplus B_1 \bigoplus E$. However, they are inequivalent with different basis functions. For type I, the decomposition is through the transformation
\begin{eqnarray*}
& & f^{A_1} = a + b + c + d, \\
& & f^{B_1} = a - b + c - d, \\
& & \boldsymbol{f}^E = (a - c,\ -b + d), 
\end{eqnarray*}
while the transformation for type-2 block is 
\begin{eqnarray*}
& & f^{A_1} = a + b + c + d, \\
& & f^{B_1} = a - b + c - d, \\
& & \boldsymbol{f}^E = (a + b - c - d,\ a - b + c - d). 
\end{eqnarray*}
The type-III block is an 8-dimensional representation of $D_4$ group and can be decomposed as $8 = A_1 \bigoplus A_2 \bigoplus B_1 \bigoplus B_2 \bigoplus E \bigoplus E$ through the transformation 
\begin{eqnarray*}
& & f^{A_1} = a + b + c + d + e + f + g + h, \\
& & f^{A_2} = a - b + c - d + e - f + g - h, \\
& & f^{B_1} = a + b - c - d + e + f - g - h, \\
& & f^{B_2} = a - b - c + d + e - f - g + h, \\
& & \boldsymbol{f}^E_1 = (a + b - e - f,\ -c - d + g - h), \\
& & \boldsymbol{f}^E_2 = (c - d - g + h,\ a - b - e + f),
\end{eqnarray*}
Feature variables that are invariant with respect to the point group symmetry can now be derived from the basis functions for the IRs. First, a set of invariants called the power spectrum can be readily obtained:
\begin{eqnarray}
	p^\Gamma_r \equiv |\boldsymbol{f}^\Gamma_r|,
\end{eqnarray}
Here $\Gamma$ denotes the IR and $r$ enumerates the multiple occurrence of $\Gamma$. However, the power spectrum alone is incomplete in the sense that it cannot distinguish different configurations that are not related by the point group operations. For instance, the power spectrum is invariant under independent rotations of each layer of the neighborhood. Formally, this spurious symmetry is due to the ambiguity of the relative ``angle" between two IRs of the same type, $\text{cos}\theta_{12}=(\boldsymbol{f}^\Gamma_{1}\cdot\boldsymbol{f}^\Gamma_{2}) / (|\boldsymbol{f}^\Gamma_{1}||\boldsymbol{f}^\Gamma_{2}|)$, which is also an invariant. Different $\theta_{12}$ correspond to distinct configurations that are not related by the point group symmetry.  The power spectrum should be supplemented by feature variables that encode the relative phases. 

A systematic approach to include all relevant invariants, including both amplitudes and relative phases, is the bispectrum method~\cite{Kondor, bartok13}. The bispectrum coefficients are triple products of IR basis defined as 
\[
	b^{\Gamma, \Gamma_1, \Gamma_2}_{r, r_1, r_2} = \sum_{\kappa, \mu, \nu} C^{\Gamma; \Gamma_1, \Gamma_2}_{\kappa, \mu, \nu} f^{\Gamma}_{r, \kappa}, f^{\Gamma_1}_{r_1, \mu} f^{\Gamma_2}_{r2, \nu}.
\]
where $C^{\Gamma; \Gamma_1, \Gamma_2}_{\kappa, \mu, \nu}$ are the Clebsch-Gordan coefficients of the point group, which are introduced to account for the different transformation properties of the three IR basis.  For a local environment of $M$ neighboring sites, the number of bispectrum coefficients scales roughly as $\mathcal{O}(M^3)$. The rather large number of bispectrum coefficients is also due to the fact that many of them are redundant. 

For practical implementation, we present a method that is modified from the bispectrum method. We introduce the reference basis $\boldsymbol{f}^\Gamma_{\rm ref}$ for each IR of the point group. Although the choice of the reference can be any of the $\boldsymbol{f}^\Gamma_r$ from the 8-site blocks, it is more desirable to build them by applying similar decomposition procedure applied to large symmetry-related blocks of $\mathcal{C}_i$ such that they are insensitive to small variations of the neighborhood~\cite{zhang22a}. We then define the ``phase" of an IR basis as the projection of its basis functions onto the reference basis: 
\begin{align}
\label{eq:relative_phase}
\eta^\Gamma_r = \langle \boldsymbol{f}^\Gamma_{r} | \boldsymbol{f}^\Gamma_{\text{ref}} \rangle 
\equiv (\boldsymbol{f}^\Gamma_{r} \cdot \boldsymbol{f}^\Gamma_{\text{ref}}) / (|\boldsymbol{f}^\Gamma_{r}| |\boldsymbol{f}^\Gamma_{\text{ref}}|), 
\end{align}
where $\boldsymbol{f}^\Gamma_r \neq \boldsymbol{f}^\Gamma_{\text{ref}}$. Importantly, the relative phases between two neighborhood IR of the same type can be obtained through the reference. 

Finally, for a complete representation, the relative phases between different IR-types should also be included in the descriptor. This can be provided simply by the bispectrum coefficients of the reference IR's ${\bm f}^{\Gamma}_{\rm ref}$ themselves. However, even this reduced set of bispectrum coefficients is highly redundant. A more practical approach is to introduce a standard form of the reference representation. To this end, we first identify the symmetry transformation $\mathcal{T}$ of the $D_4$ group such that the angle between the rotated doublet IR $\mathcal{T} {\bm f}^E_{\rm ref}$ and the unit vector $\mathbf{e}_1 = (1, 0)$ is within $[0, \pi/4]$. The resultant angle thus serves as a unique phase of the reference double IR
\begin{align}
\label{eq:T}
\eta^E_{\rm ref} = \langle \mathcal{T} \boldsymbol{f}^E_{\text{ref}} | \mathbf{e}_1 \rangle / \bigl| \boldsymbol{f}^E_{\text{ref}} \bigr|,
\end{align}
Note that the ambiguity due to the 8-fold symmetry of $D_4$ is essentially removed in this process. With the aid of $\mathcal{T}$, the phase of singlet IR in the reference is defined as
\begin{eqnarray}
	\eta^{\Gamma}_{\rm ref} = {\rm sign}(\mathcal{T} f^{\Gamma}_{\rm ref}), \qquad \Gamma = A_1, A_2, B_1, B_2.
\end{eqnarray}
 The generalized coordinates of the neighborhood $\mathcal{C}_i$ are given by $\{ G_\ell \} = \{p^\Gamma_r,\  \eta^\Gamma_r, \ \eta^\Gamma_{\rm ref}\}$.

\section{Neural network model }

\label{sec:NN}

The NN model used to predict the force is constructed and trained on PyTorch~\cite{Paszke2019, Nair2010, Barron, Paszke2017, He2015, Kingma}. A total of 6 layers is used in our NN model, with the number of neurons $45 \times 512 \times 256 \times 64 \times 16 \times 1$. The number of nodes of the input layer is determined by the number of feature variables in $\{ G_\ell \}$.  The output layer is a single neuron that gives the force. The loss function is defined as the mean square error (MSE) of the local forces:
\begin{eqnarray}
	\mathcal{L} = \frac{1}{{N}} \sum_{i=1}^{{N}} (F_i^{\rm ML} - F_i^{\rm exact})^2.
\end{eqnarray}
Three models are trained corresponding to three different electron filling fractions at 0.3, 0.4 and 0.5. For each model, 600 configuration snapshots from a $40 \times 40$ square lattice are used in the training dataset, including 200 snapshots of random configurations and 400 snapshots of equilibrium configurations at a low temperature $T = 0.1$ from the simulation. In the training, the training dataset batch size is set as 1, and 500 epoches are applied to train the model. Adam optimizer~\cite{Kingma} with an adaptive learning rate 0.001 is used for this process.

\end{document}